\documentstyle[11pt,aas2pp4]{article}
\received{February 1, 1999}
\accepted{June 9, 1999}

\newcommand{\degrees}{\mbox{$^{\circ}$}}    
\newcommand{\chisq}{\mbox{$\chi^2$ }}       
\newcommand{\NH}{\mbox{${\rm N_H}$ }}       
\newcommand{\NHunits}{\mbox{$ 10^{20}~{\rm cm}^{-2}$}}

\begin{document}

\title {HIGH RESOLUTION X-RAY IMAGING OBSERVATIONS OF TWO LOW LUMINOSITY SEYFERT GALAXIES}
\author {Kulinder Pal Singh \altaffilmark{1}}

\affil{Department of Astronomy \& Astrophysics, Tata Institute of Fundamental Research, Homi Bhabha Road, Mumbai 400005, India}
\altaffiltext{1}{singh@tifr.res.in}
\bigskip
\bigskip
\centerline{Accepted for publication in {\em MNRAS}}

\begin{abstract}

Results from observations of two {\it nearby, low X-ray luminosity}
Seyfert-type galaxies, namely, NGC 1365, and NGC 4051, using the $ROSAT$ 
High Resolution Imager (HRI) are presented.  The observations carried 
out with the aim of detecting and resolving the extra-nuclear (beyond 
the central 5\arcsec--10\arcsec) X-ray emission show evidence 
for an extended component of $\simeq$2 kpc size in NGC~1365 
and NGC~4051.  The extended component contains 56$\pm$8\%  of the 
total observed flux in the case of NGC~1365, and 21\%$\pm$6\% in 
the case of NGC~4051. 
In NGC~1365, the extended X-ray emission shows a component
aligned with the inner disk structure, another as wings or ears along
the east and west direction aligned with the inner spiral arms,
and an elongated edge-brightened structure (``chimney'') breaking 
out of the disk in the north-west direction.  
The extended soft X-ray component 
around the nucleus of NGC~4051 is co-spatial with the disk of the 
galaxy.  It also shows an elongation coincident with a 
``banana''-like feature in the north-east seen in the 6~cm radio band.  
Extensions are also seen towards the south and south-east of the nucleus.
Starburst activity driving strong winds through the disk of NGC~1365 
can account for most of the extended soft
X-ray emission in it.  In the case of NGC~4051, extended X-ray
emission probably owes its origin to both nuclear activity as well as
starburst induced activity.

The nuclear component of NGC~4051 shows strong soft X-ray variability 
with X-ray intensity changing by a factor of 2--3 on a time-scale 
of a  few  100s.   The power-spectrum of the variability has been
extended to higher frequencies compared to the previous observations,
and has now reached the true Poissonian noise level due to the source.  
At lower frequencies the power spectrum is best characterised by a
power-law and a Gaussian.  The power-law slope of $\simeq$-1.8 is 
consistent with the previous low-energy observations
with $EXOSAT$, and the presence of a Gaussian feature signifies the
persistence of a quasi-periodic oscillation that was also seen 
earlier with $EXOSAT$.
During its low intensity and low variability state, however, the 
extended component of soft X-ray emission dominates
the flux in NGC~4051.

A list of X-ray sources, some of which are new, detected in the field 
of view of the HRI during the observations, is also presented.

\end{abstract}

\normalsize

\section{INTRODUCTION}

X-ray emission from active galaxies is generally dominated by  
luminous unresolved nuclear sources.  In other bands, like radio 
continuum and optical line emission from ionised gas, extended
emission is seen around the nuclei of Seyfert-type galaxies.  
These extra-nuclear emissions, as distinct from the emission from
the disks of galaxies, are quite often anisotropic or collimated in 
nature suggesting an outflow of gas from the nucleus in the form
of jets or starburst-driven superwinds.  
Since these processes can also lead to X-ray emission, 
spatially extended circum-nuclear X-ray emission may also
be quite common in these galaxies.    
Observationally, its occurrence in several Seyfert galaxies with 
good high resolution observations has been pointed out by 
Elvis et al. (1990) and Wilson (1994), and confirmed in five
Seyferts viz., NGC~1068, NGC~1566, NGC~2110, NGC~2992 and
NGC~4151 (see Weaver et al. 1995).  The extent of the circum-nuclear
X-ray emission in these Seyfert ranges from $\sim$7\arcsec~to
$\sim$30\arcsec, going up to 90\arcsec~in NGC~1068 which includes
the starburst regions (Wilson et al. 1992).
The intensity of the extended circum-nuclear component amounts 
to $\sim$ 10\% $-$ 30\% of the total soft X-ray flux in these objects.

In principle, the extended X-ray emission could arise due to a 
number of reasons.  For example,   thermal 
emission from hot interstellar gas, emission from starburst regions, 
synchrotron radiation or inverse Compton scattering by relativistic
electrons, and electron-scattered nuclear radiation all could lead to
such an emission.   
It is, however, believed that the inner extended ($\sim$ 10\arcsec) 
X-ray emission is most probably related to the active nucleus because 
of its abnormal brightness, whereas the emission on the arcmin
scale could be due to starburst regions with a large number of 
binaries and supernova remnants (Elvis et al. 1990; Wilson et al. 1992).
Another favoured explanation for the extra-nuclear component 
is thermal emission from hot gas in pressure equilibrium with 
the optical narrow-line gas, although other possibilities cannot be 
ruled out at present. 

$ROSAT$ with its low-scatter mirror and High Resolution Imager (HRI), 
has the unique capability to resolve extended X-ray components beyond 
the central 5\arcsec~of the point-like nucleus of an active galaxy 
(also known as the active galactic nucleus or AGN) as has been 
demonstrated in some of the cases mentioned above.  Furthermore,
the weak extra-nuclear emission is best resolved in nearby and 
low-luminosity AGN's, where the nuclear component is less dominant and is 
further suppressed by high absorption  intrinsic
to the nuclear region of these objects, thus reducing 
the contribution from the point spread function (PSF)
to the extra-nuclear regions.  With this consideration in mind 
I have observed two {\it low luminosity, nearby (distance $\leq$ 20 Mpc) 
and hard X-ray selected} Seyfert galaxies, viz., NGC 1365 and NGC 4051
with the $ROSAT$ HRI.
The Galactic column density due to the interstellar medium towards them is
very low, which is also an important consideration while looking
for an extended X-ray component.   

X-ray variability is another characteristic of an AGN that holds clues
to the physics of the central regions.  $ROSAT$ with its high sensitivity
in soft X-rays can probe variations at much smaller time scales than
before. Of the two galaxies targeted for observations, NGC~4051 is known
to be highly variable on all time scales observed so far.  Extending
the study of its variability to short time-scales could be very useful,
as is indeed found in the present observations.

In this paper, I report $ROSAT$ HRI observations of NGC~1365 and
NGC~4051 and their spatial and timing analysis.  
The paper is organized as follows.
In \S2, I review a few of the basic parameters  of the target
galaxies, followed by observational details in \S3.
In \S4, I present the  results from a detailed spatial analysis
of the X-ray images (corrected for the instrumental wobble),  
their comparison with other wavebands, 
and a temporal analysis of short term variability seen in NGC~4051.  
The results are discussed in \S5, followed by conclusions in \S6.
  
\section {The Sample}

Some of the basic parameters of the two Seyfert galaxies are listed 
in Table 1.  These include their distance, B magnitude, 
hard X-ray luminosity in the 2 - 10 keV band, the N$_{\rm H}$ 
from 21 cm measurements (Elvis, Lockman \& Wilkes 1989; Dickey
\& Lockman 1990), soft X-ray extent, soft X-ray flux and luminosity 
in the 0.2 - 2.0 keV band from the present observations.  Other
important characteristics of these galaxies are given below.

\subsection {NGC~1365}

NGC 1365 is a luminous, dusty, barred spiral galaxy belonging to the
Fornax cluster of galaxies.  It has been classified as type 
SBb(s)I (Sandage \& Tammann 1981).  Measurements of 37 Cepheids 
in NGC~1365 by Madore et al. (1998) give a distance of 18.6$\pm$1.9 Mpc. 
It has large extinction and several bright H$\alpha$ spots near 
its nuclear region. The optical spectrum of its nucleus shows both 
broad and narrow H$\alpha$ emission and forbidden narrow-line emission, 
and therefore, it can be called a Narrow Emission Line Galaxy 
(NELG) (Veron et al. 1980) 
or a Seyfert type 1.5, although Veron \& Veron (1989) (VV89) list it as
a type 1 Seyfert.  It shows a small radio jet and a circum-nuclear ring
with a number of non-thermal continuum radio sources in it
(Sandqvist et al.  1995).  A conical outflow of
high excitation gas, seen in [O III], has been found aligned with the small
radio jet (Kristen et al. 1997).  The Galactic \NH~ towards this galaxy is
1.35$\times$\NHunits~(Dickey \& Lockman 1990).
In soft and hard X-rays, it is an order of magnitude less luminous than 
most type 1 Seyferts.  NGC~1365 was  perhaps the first Seyfert for 
which an extended soft X-ray emission was suspected (Maccacaro et al.1982).  
The X-ray spectrum  of NGC~1365 has been found to contain a hard power-law
component and a soft thermal component based on $ROSAT$ observations 
by Turner, Urry, \& Mushotzky (1993) and Komossa \& Schulz (1998), and
$ASCA$ observations by Iyomoto et al. (1997).  A strong line emission
component due to iron is also detected in the $ASCA$ observations
by Iyomoto et al. An analysis of the $ROSAT$ HRI observations
of NGC~1365 has been presented by Komossa \& Schulz (1998) in which
they pointed out the detection of a bright variable source 
southwest of NGC~1365.

\subsection {NGC~4051}

NGC 4051 is an early type spiral, SB (Paturel et al. 1989).  It
has an optically bright and active nucleus that shows both broad and
narrow permitted line emission and forbidden narrow-line emission. 
It has been classified as Seyfert type 1.5 by Dahari \& De Robertis (1988).
The Galactic \NH~towards this galaxy is 1.3$\times$\NHunits~
(Elvis et al.  1989).  Among the nearest (distance = 13.8 Mpc) 
Seyfert galaxies,  it has the lowest X-ray luminosity of all the 
galaxies of this type.  A triangular region of ionised gas seen in
[O III] line emission has been detected out to a distance of $\sim$400 pc
from its centre by Christopoulou et al. (1997).  Anisotropic radio
emission from this galaxy has been studied by Baum et al. (1993) and
Kukula et al. (1995).  Three components of radio emission -- nuclear,
extra-nuclear, and galactic disk have been seen in radio maps of different
resolutions in 6cm and 21cm continuum bands (Baum et al. 1993). 
The [O III] line emission region is co-spatial with the extended radio 
emission (Christopoulou et al. 1997).

It is a highly variable X-ray source (see Papadakis \& Lawrence 1995),
and its spectrum has been measured with many instruments -- 
$HEAO$-1 (Marshall et al. 1983), $EXOSAT$ (Turner \& Pounds 1989), 
$Ginga$  (Matsuoka et al. 1990; Nandra \& Pounds 1994), 
$ROSAT$ (McHardy et al.  1995; Komossa \& Fink 1997), 
simultaneous $ROSAT$ and $Ginga$ (Pounds et al. 1994), 
and $ASCA$ (Mihara et al 1994). 
Its broad-band X-ray spectrum can be described by a power-law 
with photon index in the range of 1.6 to 2.3, 
and modified by the presence of a warm absorber (Pounds et al. 1994;
Mihara et al. 1994).
The steepness of the power-law is usually positively correlated with the
brightness of the X-ray source (Matsuoka et al. 1990).
A steeper low energy ($<$1 keV) spectral component (``soft excess'') 
due to a  black-body or hot thermal gas at a few million degrees K  
has also been seen in many observations (Pounds et al. 1994; 
Mihara et al. 1994).  The soft excess is, however, usually seen 
in X-ray high states analysed by Pounds et al. (1994) and 
Mihara et al. (1994)  but not in X-ray low states analysed by 
McHardy et al.  (1995) and  Komossa \& Fink (1997).  However, in a 
very low X-ray state observation, a steep low energy component 
with photon index of 3.0$^{+0.2}_{-0.3}$ has been 
reported by Guianazzi et al. (1998).  During this very low state,
both the hard X-ray power-law component as well as the soft 
component were significantly lower than in the previously reported
observations.

\section {OBSERVATIONS}
$ROSAT$ HRI (Truemper 1983,  Pfeffermann et al. 1987) observations 
of NGC~1365 and NGC~4051 were proposed by me for a total of 
20,000s and 10,000s, respectively.  These exposure times
were realised in observations carried out in 1994 and 1995.   
NGC~4051 was observed in 1994, whereas the observations 
of NGC~1365 were split-up over the two years and many satellite orbits. 
Further details of these observations  are given in Table 2.  
In the case of NGC~1365, data from only two of the longest observations 
were used for further analysis below.
The known positions of the centres of the galaxies were targeted
at the on-axis positions of the HRI, and X-ray peaks were detected 
within 2\arcsec~of these positions. 

Data were recorded over several satellite orbital intervals (OBIs). 
X-ray images were made using data from all the OBIs  in the case of
NGC~4051, and most of the OBIs in the case of NGC~1365, after correcting
for the residual errors from aspect corrections and due to wobble of the
satellite.  Details of the
analysis and results are given in the following section.

\section {ANALYSIS AND RESULTS}

\subsection {Count Rate and X-ray Flux Measurements}

The total counts for the targeted sources were extracted from the X-ray 
images using a circle of radius 22\arcsec~for NGC~1365 and 30\arcsec~for
NGC~4051, centred on the source. Background counts were obtained from an
annulus 50\arcsec~wide outside a radius of 80\arcsec~from the source 
centre, and subtracted from the total source counts after scaling
the background to the same number of pixels as in the source.
The mean count rates thus derived for sources centred on the nuclei
of NGC~1365 and NGC~4051 are listed in Table 2.

The conversion of the observed count rate to the X-ray flux requires
the knowledge of the X-ray spectrum and the characteristics of
the telescope and detector.  For NGC~1365, spectral measurements
have been provided by Turner et al. 1993 using the  $ROSAT$ 
Position Sensitive Proportional Counter (PSPC), and by Iyomoto et 
al (1997) using the  $ASCA$ Solid-State Imaging Spectrometer (SIS).
The $ROSAT$ PSPC  measurements over the 0.1--2.0 keV energy range suggest
the presence of a power-law with energy index ($\alpha$) of 1.22
and a thermal component (Raymond-Smith plasma model) with kT=0.55 keV
(Turner et al. 1993) with the two components being equally strong
at 1 keV (ratio of about 1:0.90).  Using this spectral model and
the $ROSAT$ HRI characteristics I estimated the X-ray flux in the
$ROSAT$ HRI band of 0.1--2.0 keV using the `hxflux' program in the
PROS software package.  The absorption in the line of sight to the
source was kept fixed at the 21-cm value of 1.35$\times$\NHunits~and the
abundances used were as given by Morrison \& McCammon (1983).
The unabsorbed X-ray flux from NGC~1365 thus obtained  is
1.08$\times$10$^{-12}$ ergs cm$^{-2}$ s$^{-1}$, with the corresponding
X-ray luminosity being 4.5$\times$10$^{40}$ ergs s$^{-1}$. These values
are given in Table 1. (A similar result is obtained by using the 
spectral model of Komossa \& Schulz (1998).) 
The $ASCA$ SIS measurements over the 0.6--7.0 keV band also suggest 
the presence of both a  power-law and a thermal component, as well 
as a strong line emission due to iron (Iyomoto et al. 1997).  
The energy index of the power-law in these
measurements is, however, very flat (-0.2), and the temperature
of the Raymond-Smith plasma model is 0.85 keV with the abundance being
only 20\% of the solar value.  Use of this spectral model for the
present $ROSAT$ HRI observations gives a flux that is $\sim$24\% less than 
the value given above in the 0.1--2.0 keV energy range.

There have been many spectral measurements of the highly
variable X-ray source in NGC~4051 as mentioned in \S2.2.
The spectral parameters can vary quite rapidly in response to
the intensity variations, making the task of converting total
HRI count rate to flux rather difficult.   Presently, it is not 
clear what spectral parameters can be attributed to the extended
soft X-ray component in NGC~4051.  However, for the purpose of 
estimating the flux associated with such a component, I assume 
that the total soft X-ray flux can be characterised 
as a power-law with $\alpha$ assumed to lie in the range of 
0.6 to 2.0, as indicated by the various measurements.  
The absorption in the line of sight to the source was kept fixed 
at the 21-cm value of 1.3$\times$\NHunits~with the elemental
abundances as given by Morrison \& McCammon (1983).  
The total count rate  corresponds to the observed (absorbed) flux
of $\simeq$ 3.20$\times$10$^{-11}$ ergs cm$^{-2}$ s$^{-1}$ for 
$\alpha$ = 0.6--1.0 in the 0.1--2.0 keV energy range.  
For $\alpha$ = 1.5, 2.0 the corresponding flux values are
3.0, 2.8$\times$10$^{-11}$ ergs cm$^{-2}$ s$^{-1}$, respectively. 
The unabsorbed source flux in the same energy is, however, 
very susceptible to the value of $\alpha$;  the total flux
values being 4.3, 4.75, 5.74 or 
7.2$\times$10$^{-11}$ ergs cm$^{-2}$ s$^{-1}$ corresponding
to $\alpha$=0.6, 1.0, 1.5, or 2.0 respectively.
Adopting a value of 1 for  $\alpha$, the X-ray luminosity of the 
source is 1.08$\times$10$^{42}$ ergs s$^{-1}$.

\subsection{Other X-ray Sources in the Field of View}

X-ray emission was also detected from several new objects.  These
sources are listed in Table 3.  These sources were found using the
source detection programs in the XIMAGE and PROS software packages.
Only sources $\geq$ 3$\sigma$ above background noise are shown,
along with the count rates determined from a box size of 
8\arcsec$\times$8\arcsec~around the source.
Background was determined from a box size of 32\arcsec$\times$32\arcsec,
but excluding the central source box and has been subtracted from
the count rates shown in Table 3. 
Many of these sources are variable.  Even though the exposure times in 1994
and 1995 were almost equal, some of these were detected either in 1994 or
in 1995, but not in both years, and therefore termed as transients.  
This information is provided against each source in Table 3.

In a previous study based on PSPC observations in 1991-92, Turner et al.
(1993) had detected a number of sources in the field surrounding NGC~1365.
Two of those sources have been detected in the present observations, and
one of them has been resolved into two separate sources (source numbers 
1,2,3 in Table 3).   Sources 5 and 6
are coincident with a strong source detected with the Einstein Observatory.
Source number 7 is  coincident with a clusters of galaxies.
Because of the better position determination in the present observations
compared to that given in Turner et al. (1993), I have 
looked for the optical counterparts of these sources using the Digitised
Sky Survey data and the USNO catalogue provided on-line by ESO via 
Skycat.  Based on this, I provide the information about the possible
optical counterparts of these sources in Table 3.

\subsection {Resolving the Spatial Components}

The knowledge of the pointing position of the telescope as a function of
time is required to be known to a very high accuracy ($<$ 4\arcsec)
to be able to fully exploit the intrinsic spatial resolution 
(4\arcsec~-- 5\arcsec~FWHM) of the $ROSAT$ HRI for detecting extended X-ray
emission within 10\arcsec~-- 20\arcsec~of bright, unresolved point sources
(David et al. 1993).  The errors in determining the pointing position
(aspect errors) can arise from either the $ROSAT$ wobble (introduced
to protect the HRI and reduce shadows due to mechanical structures)
or errors in the re-acquisition of the guide stars.  Aspect errors can 
lead to artificial elongated structures around point sources or blurring 
on the scale of 10\arcsec~(David et al.1993).  Corrections for these
errors have been provided by the analysis of Morse (1994).  Recently,
Harris et al. (1998) have provided a set of scripts
under the IRAF/PROS package to minimise these effects.
These routines correct the aspect for 
a) residual errors induced by the wobble of the satellite, and 
b) errors due to re-acquisition of the guide stars at the start of 
each OBI in a multi-OBI observations.
Errors due to wobbling are corrected for by dividing the data in
different phase bins of the wobble period of 402s. 
Harris et al. (1998) give examples showing the improvement
in the HRI images after applying the above corrections to aspect,
which can vary from observation to observation.

Observations of NGC~4051 were done in 2 OBIs, whereas the 
observations of NGC~1365 in 6 OBIs in 1994 and 5 OBIs in 1995. The
effect of aspect errors could be seen easily in the case of the strong
source in NGC~4051 resulting in displaced images in different OBIs.
I have used the  routines provided by Harris et al. (1998) to find 
the aspect solutions, and restacked the images for each OBI to produce
the corrected image for each source using all available data.
In the case of NGC~4051, I was able to divide the data in 10 phase bins
for finding the centroid of the image for each phase bin of the wobble
period in an OBI while using an aperture factor of just 3 times 
the $\sigma$ of the PSF of the HRI for accepting the source photons, 
and then restacking the images at a position closest to the optical 
position of the source.  The centroids of the images in different
phase bins were found to wander around by as much as 11\arcsec
in the y-axis (declination) and $\simeq$3\arcsec in the
x-axis.  The entire data were used and corrected for errors due to 
wobble as well as due to the re-acquisition of the guide stars.
The resulting X-ray image of NGC~4051 is  shown in Figure 1(a) after
using Gaussian smoothing with $\sigma$=2\arcsec, and in Figure 1(b)
after using  Gaussian smoothing with $\sigma$=5\arcsec to show both 
the small scale and larger scale structures in X-ray emission.
The $uncorrected$ image, after using Gaussian smoothing with
$\sigma$=2\arcsec, is shown in Figure 1(c) for comparison with
Fig. 1(a).  An elongation of the inner contours 
(radius $\leq$10\arcsec) along the south easterly direction can be 
seen in Fig 1(c), and is mostly due to the aspect errors in the data.
The peak brightness is $\sim$12\% lower than that in Fig. 1(a).
These images (Figs. 1(a),(b),(c)) of surface brightness have been 
shown as contour plots with contours plotted at 
0.1, 0.15, 0.2, 0.3, 0.5, 1, 2, 5, 10, 20, 30, 50, 70, 90 and 95\% of 
the peak brightness level.
The mean background level in Fig.1(a) is $\sim$0.015$\pm$0.003 
counts/pixel and that in Fig.1(b) is $\sim$0.005$\pm$0.001 counts/pixel.
The lowest contour plotted in the figures 1(a) and 1(b) is
of $\geq$ 2$\sigma$ significance above background.
A number of anisotropically extended components can be seen in these
contour maps towards the north-east, south and south-west of the nucleus.

In the case of the much weaker source in NGC~1365, the aperture factor
had to be increased to 10$\sigma$.  Besides, the 1995 data were
not divided into any phase bins and the different OBIs were corrected
for errors associated with the re-acquisition of guide stars only.  
The 1994 data of NGC~1365 did, however, require division of the data into
2 phase bins to get rid of the residual wobble effects as well as errors
due to the re-acquisition of guide stars.  This resulted in some
loss of data, and the effective exposure time for the 1994 data used
for analysis was thus reduced to 5904.4s.
The resulting images from 1994 and 1995 data were then added 
to produce the final X-ray images of NGC~1365 shown in Figures 2(a) and
2(b).  As in Figures 1(a) \& (b), Gaussian smoothing with $\sigma$=2\arcsec~
has been used in Fig. 2(a) and with $\sigma$=5\arcsec~in Fig. 2(b).
The effective exposure time for the images shown in Figs. 2(a) 
and 2(b) is 15666.4s.  Contours are shown plotted at 5, 8, 12, 20, 30,
50, 70, 90 and 95\% of the peak brightness level in Fig 2(a), and
4, 6, 10, 20, 30, 50, 70, 90 and 95\%  of the peak brightness level 
in Fig 2(b).  The mean background level in Fig.2(a) is 
$\sim$0.008$\pm$0.007 counts/pixel and that in Fig.2(b) is 
$\sim$0.005$\pm$0.002 counts/pixel.
The lowset contour plotted in the figures 2(a) and 2(b) is
of $\simeq$ 2$\sigma$ significance above background.
Several anisotropically extended components can be seen around the 
central amorphously extended emission, particularly in the 
east-west directions.

The azimuthally averaged radial brightness distributions of the
X-ray sources in the galaxies were extracted from the corrected
images and are shown in Figures 3 and 4, respectively for NGC~1365 and
NGC~4051.  The azimuthally averaged point source radial profile
given by David et al. (1993) was computed and scaled to match the
inner region ($\leq$ 5\arcsec) of each X-ray source, and is shown as
a solid curve in Figs. 3 and 4.  Excess X-ray emission is seen clearly
in the two sources at extents between 6\arcsec~ and 20\arcsec.  
A circularly symmetric point spread function (PSF) image was also created 
using the azimuthally averaged point source radial profile
given by David et al. (1993).  The excess due to the extended emission
component was determined by binning the source and the scaled PSF
images radially over identical regions and then subtracting the 
contributions from the PSF.  The extended emission component found
thus in the case of NGC~4051 is 0.229$\pm$0.014 counts s$^{-1}$, 
amounting to 21\%$\pm$6\% of the total observed count rate and 
extending to a radius of $\simeq$30\arcsec.
About 68\% of this extended component is in the radial bins from 
6\arcsec~to 20\arcsec.  In the case of NGC~1365, the excess over the
PSF or the extended
component was determined similarly from the re-stacked images from
1994 and 1995 observations.  The total average count rate for the source
was found to be 0.022$\pm$0.0013 counts s$^{-1}$, of which
0.0126$\pm$0.0007 counts s$^{-1}$ were due to the extended component
in the region encompassed by radii of 6\arcsec~to 22\arcsec~centred
on the source.  This amounts to 56$\pm$8\% of the total observed 
X-ray count rate (not corrected for absorption) being
in the circum-nuclear component of NGC~1365. 

\subsection {Comparison with Optical and Radio Emission}

Anisotropic emission from circum-nuclear regions of these galaxies 
has also been seen at other wavebands, e.g., radio continuum 
and [O III] 5007\AA~line emission.  A comparison of these extended 
emissions in various wavebands is presented below.

\subsubsection {Extended Components in NGC~1365}

High quality optical images of NGC~1365 in the B-band and [O III]
have been published by Kristen et al. (1997) (see their Figs. 1 and 2(b))
and were obtained from them.  Using positions of stars
(or star-like objects) in the field from Skycat program, I created 
the co-ordinate
reference system for their B-band image and overlaid the X-ray contours 
from Fig. 2(a) on the B-band image after resampling the X-ray image.
This overlay is shown in Figure 5.  The same co-ordinate system was 
also used for the [O III] line image and a similar comparison
was carried out.

From Fig. 5 one can see that the extended X-ray emission shows the 
following different features --
(a) a component aligned with the inner disk structure along the
north-east to the south-west directions, 
(b) wings or ears along the east and west direction aligned with 
the inner spiral arm structures, and 
(c) an elongated structure (chimney) in the north-west direction that 
appears to protrude through the disk of the galaxy. 
This structure shows an edge brightening towards its northern end,
and is aligned with a channel through the disk as seen in the blue 
colour image in that direction.
No X-ray features seem to be related to the [O III] conical 
feature or the small radio jet reported in the literature.
The component (a) above is also aligned with the circum-nuclear
elliptical ring seen in the radio continuum by Sandqvist et al. (1995),
and is of similar extent ($\sim$45\arcsec).  The radio ring has a 
channel of low intensity, just like in the optical, along the 
component (c) in X-rays.

\subsubsection {Extended Components in NGC~4051}

Good quality images of NGC~4051 in the continuum of V-band and pure
[O III] line emission published by Christopoulou et al. (1997) (see their
Fig. 1) were obtained from them.  Because of the lack of any stars
in the field, the optical images were re-scaled, shifted, and trimmed 
to align the nuclear region with the X-ray image as well as possible.
A residual mis-alignment of $\sim$2\arcsec~ could have remained, however. 
An overlay of the X-ray contours from Fig. 1(a) on the V-band image of
is shown in Figure 6.  We have also created an X-ray image of the
anisotropic X-ray component by subtracting a model image of the 
point spread
function scaled to the central (inner to 4\arcsec) intensity of NGC~4051.
X-ray intensity contours from this image have also been overlaid on the
V-band image and this is shown in Figure 7.
Similarly, an overlay of the X-ray contours from Fig. 1(a) on 
[O III] line emission regions is shown in Figure 8.

The extended component of X-ray emission (Figs. 6 \& 7) 
appears to be associated with the disk of the galaxy and 
coincident with it.   X-ray emission also
shows an elongation along the banana-like feature in the north-east 
seen in the 6~cm radio band (Baum et al. 1993; Kukula et al. 1995) which 
is also aligned with the [O III] emission cone (Fig. 8).  
Extended X-ray components are also seen (a) towards the south-east 
of the nucleus, and (b) towards the south.  The component towards 
the south appears stronger in the high resolution X-ray maps 
(Figs. 1(a) \& 6), whereas the component towards  the south-east is
stronger in the low resolution X-ray map of Fig 1(b). 

\subsection {Variability and Power Spectrum of NGC~4051}

X-ray source counts from a region (radius = 1\arcmin) centred on the
source peak were extracted and examined for source variability.  
An X-ray intensity curve was derived after binning the data every 32~s, 
and is shown in Figure 9 as a function of time. 
The source strength varies from 
$\sim$ 0.5 counts s$^{-1}$ to about 1.8 counts s$^{-1}$, with the 
average intensity being around 1 count s$^{-1}$.  The time-scale for
this variation seems to be as small as 200s, and is significantly
smaller than the previous time-scales observed with the less sensitive
low energy experiment aboard $EXOSAT$.  The observed
source intensity in soft X-rays indicates that the source was in a
high state during these observations in 1994.  To estimate the
power spectrum of this light curve, I created two light curves:
(a) using data binned every 16s, and (b) using data binned every 32s.
We used the `powerspec' program in FTOOLS to estimate the fractional
power as a function of frequency. The power spectra were estimated for
128 bins per interval for both light curves, and then averaged.  
The averaged power spectra thus obtained are plotted in units of 
(rms)$^2$ Hz$^{-1}$ in Figure 10.
The power spectrum based on finer binning of 16s shown in the upper 
panel extends to the highest frequency observed so far.  
The constancy of the spectrum for frequencies higher than 0.005 Hz shows 
the onset of the Poisson noise in the source.
The spectral power increases towards lower frequencies, between
0.005 and 0.0005 Hz.  A constant plus a power-law fitted to the 
power spectrum shown in the upper panel in Figure 10 gives
a power-law index = -2.1$\pm$0.4.
The power spectrum based on time binning of 32s,
shown in the lower panel of Fig.10, is more sensitive to the lower
frequencies and is used to further investigate the nature of excess power
at low frequencies.  A constant plus a power-law fitted to the 
power spectrum shown in the lower panel in Figure 10 gives
a power-law index = -1.8$\pm$0.3.  This model is shown as a dotted line 
in the lower panel of Fig.10 and has $\chisq$=52.45 (for 29 degrees 
of freedom (dof)).
An excess power above the dotted line can be seen at frequencies 
lower than 0.002 Hz.  We, therefore, fitted
the power spectrum  with a constant+power-law+Gaussian model.  
The best fit model has $\chisq$=45.06 (for 26 dof) and is shown as
a solid line in the lower panel.  Based on the F-statistics the
improvement in the fit is significant at more than 99\% confidence level. 
The power-law index is now -1.8$\pm$0.5, and the Gaussian is centred at
0.0011$\pm$0.0004Hz and has a width of 0.00042$\pm$0.00022Hz.

\section {DISCUSSION}

Based on high resolution X-ray observations with the $ROSAT$ HRI, extended 
and anisotropic soft X-ray emission components have been detected around 
the nucleus of both NGC~1365 and NGC~4051.  Although the presence of 
a soft extended component in NGC~1365 had previously been indicated 
by IPC observations, the PSF of the IPC was, however, not best suited 
for this problem.  In addition, the presence of several nearby
sources detected with the $ROSAT$ PSPC (Turner et al. 1993), some of
them variable, might have significantly confused the IPC measurements. 
The origin of this extended emission and its relation to other 
wavebands is discussed below, and is followed by a discussion of 
rapid X-ray variability in NGC~4051.  

\subsection{Extra-nuclear X-ray Emission}

The circum-nuclear X-ray emission, over and above the strong nuclear
component, is found to extend to a radius of $\sim$2 kpc in both
NGC~1365 and NGC~4051.
The extent of this extra-nuclear X-ray emission is much larger 
than the extent of emission line gas in these galaxies.  
In NGC 1365, none of the extended features is coincident with 
the emission line gas.  The X-ray extent is, however, 
consistent with the  size of the radio disk, and also with the
starburst regions in the optical disk.  
Similarly, the X-ray extent in NGC 4051 is also consistent with 
the size of its radio disk.  
An extended X-ray emission feature in NGC~4051 appears related 
to the jet-like feature seen in radio.
A region of X-ray emission resembling a ``chimney'' and suggesting 
a breakout of superwind from the starburst regions in the disk 
is clearly visible in NGC~1365.
The bulk of the extended X-ray emission is, therefore, very likely to 
come from these starburst regions.  

The detection threshold, corresponding to signal-to-noise ratio of 2.5
for a point source in the present observations, 
is $\sim$4$\times$10$^{39}$ ergs s$^{-1}$ in the case of NGC~1365, 
and $\sim$2.2$\times$10$^{39}$ ergs s$^{-1}$ in the case of NGC~4051.
This threshold is above the luminosity of most of the known types of
individual sources (for example, X-ray binaries, supernova 
remnants etc.) in Local Group galaxies, although a few sources
have been reported to come close to this limit or even exceed it,
as for example in NGC 1365 itself (Iyomoto et al. 1997).   
X-ray sources which are below the detection threshold, and
thus could not be detected individually, can 
collectively contribute to the extended emission.  
The X-ray luminosity of 2.5$\times$10$^{40}$ ergs s$^{-1}$ for
the extended component in NGC~1365 is comparable
to that seen in NGC~253 and other starburst galaxies (Fabbiano 1988).
Thus, about 1000 luminous (L$_x$=10$^{37}$ ergs s$^{-1}$)
X-ray binaries and a similar number of supernova remnants can 
account for the extended X-ray emission in the disk of NGC~1365.
This is not unreasonable for a supernova rate of $\geq$ 0.1 yr$^{-1}$,
seen in starburst galaxies, over 10$^7$-10$^8$ years of a starburst 
age and a lifetime of 10$^3$--10$^4$ years for the
bright phase of X-ray emission from supernova remnants.
The ``chimney'' seen in  NGC~1365, however, suggests a superwind
activity from starburst driven hot winds,  pointing to 
disk-halo interaction in NGC~1365 and the presence of a 
diffuse X-ray component. Superwind models of Suchkov et al. (1994)
predict the formation of such ``chimney'' like structures with 
edge-brightening due to shocked gas at the top heated by superwinds
from starbursts in the disks. An edge-brightened feature is indeed
present in the ``chimney'' feature in NGC~1365.  A thermal component with 
the temperature of a few million degrees K as predicted in the
simulations of Suchkov et al. also appears to be present in the spectrum of
NGC~1365 (see \S 4.1) and could in fact be associated 
with the extended soft X-ray emission observed.  A detailed discussion
of the contribution of starburst driven superwinds to the X-ray luminosity
of NGC~1365 has been given by Komossa \& Schulz (1998). 
They have pointed out that the extremely large 
infra-red (IR) luminosity, log L(25--60$\mu$m)=44.87, of NGC~1365
measured with $IRAS$, is in considerable excess of what can be
expected from a pure AGN based on correlations between hard X-ray 
and infra-red luminosities (Ward et al 1988), and could plausibly arise
in star formation.  Such large IR luminosity if assumed to come
from a starburst would predict a supernova rate of between 0.01 and
1 yr$^{-1}$ depending on the initial mass function of the starburst and
the upper and lower mass limits in the starburst  models
of Gehrz et al. (1983).
They conclude that a supernova rate of 0.015 yr$^{-1}$ is consistent 
both with starburst models of Gehrz et al. (1983) and with the observed
soft X-ray luminosity of NGC~1365.  Their scenario also requires 
that 90\% to 99\% of the H$\alpha$ emitting gas in the starburst region
be obscured.

The extended X-ray emission component in NGC~4051 is an order of 
magnitude more luminous than in NGC~1365, but comparable to 
a similar component seen in NGC~1808 -- a superwind powered 
starburst (Dahlem, Hartner, \& Junkes 1994).  It is, therefore, more
difficult to explain by invoking unresolved X-ray emission
from discrete X-ray sources alone.  If powered by superwinds from a
starburst, then the starburst is required to be much stronger than
in the case of NGC~1365 to explain this emission. 
However, unlike in NGC~1808, there is no clear evidence for a 
strong starburst activity in the circumnuclear region of NGC~4051, 
although observations of: 
(a) widespread diffuse H$\alpha$ emission within 15\arcsec
radius from the nucleus, and compact HII regions just beyond that,
reported by Evans et al. (1996);  
(b) strong ``bar'' strength (see Martinet \& Friedli 1997, however), and 
(c) extended feature in radio (``disk'' component reported by Baum et al.
and attributed by them to superwind from a circum-nuclear starburst) 
and X-ray emission being co-spatial, indicate the presence of a 
starburst-like activity in NGC~4051 at some level.  Therefore, some
contribution to the extended X-ray emission from starburst related
activity can not be ruled out, although it may be difficult to quantify
at the moment.
The shape and extent of  [O III] emission line gas and the asymmetric 
profile of the [O III] 5007~\AA~line in and around the nucleus of 
NGC~4051 favour models with centrally driven radial 
outflow of gas (Christopoulou et al. 1997; Veilleux 1991), thus indicating
nuclear activity as possible contributor to the extended X-ray 
region in NGC~4051.  There is, however, a lack of detailed spatial 
correlation 
of extended X-ray emission with the [O III] emission line gas
in both galaxies which argues against the the extra-nuclear soft X-ray
component being thermal emission from hot gas in pressure equilibrium 
with the optical narrow-line gas.   The origin of the extended 
X-ray emission in NGC~4051 is, therefore, not very clear, and could
be due both to nuclear activity as well as starburst induced activity.

In NGC 4051, the unabsorbed soft (0.1--2.0 keV) X-ray flux from the 
extended component (21\%$\pm$6\% of the total count rate) 
corresponds to a value of 
(0.7 -- 1.0)$\times$10$^{-11}$ ergs cm$^{-2}$ s$^{-1}$.   
This is not inconsistent with the value of 
(0.4--1.0)$\times$10$^{-11}$ ergs cm$^{-2}$ s$^{-1}$ observed
by $BeppoSAX$ recently when NGC~4051 presumably went into its lowest
observed state (see upper panel of Fig. 2 in Guainazzi et al. 1998).
The spectral index at low energies ($<$ 4 keV) measured by 
Guainazzi et al. is very steep ($\alpha$ = 2.0) and a thermal origin 
has been suggested by Guainazzi et al.  In  such cases, the exact 
matching of soft X-ray fluxes is critically dependent on the 
spectral parameter characterisation and the intensity level,  
however, based on the present comparison, it is quite possible that 
the $BeppoSAX$ soft X-ray measurements indeed saw mostly the 
extended disk component.  
This is further corroborated by the near absence of soft X-ray 
variability when the source exhibits the lowest count rate in the
present observations (see the left side of the lowest panel in Fig. 9).
During this stage the extended component is nearly 50\% of the 
total count rate.

\subsection {Rapid X-ray Variability of NGC~4051}

In an AGN the X-ray source is powered by accretion onto a putative
supermassive black hole (SMBH). 
If the highly variable component of the X-ray emission in NGC~4051 comes 
from outside 3 Schwarzschild radii of the SMBH, then the shortest
observed variability time scale of 200s implies an upper limit
of 4.4$\times$10$^6$ M$_{\odot}$ for the mass of the black-hole.
A similar upper limit on the mass of the black hole was given 
by Mihara et al. (1994).

The knowledge of the form of the power-spectrum of variability
can help us investigate physical models of X-ray emission
in the source. 
In a detailed study of soft and hard X-ray variability in NGC~4051
based on long $EXOSAT$ observations, Papadakis \& Lawrence (1995) found
that at lower energies ($<$2 keV)
the power spectrum is steeper than at higher energies 
(2 -- 10 keV)  i.e., power-law slope = -1.85 versus -1.46 in the frequency
range of 10$^{-5}$Hz to 10$^{-3}$Hz, and constant at higher frequencies.
It should be noted, however that $EXOSAT$ was sensitive mostly
to the low-frequencies (with bin time=300s) and covered the
frequency range between 10$^{-5}$Hz and 4$\times$10$^{-3}$Hz.
Subsequent to $EXOSAT$ observations,  hard X-ray observations
by Matsuoka et al. (1990) using $Ginga$ showed that in hard X-rays 
the variability time scale can be as short as 200s, thus extending 
the power spectrum to higher frequencies without any noticeable 
flattening upto 0.005 Hz.
The present soft X-ray observations with  $ROSAT$ are more sensitive 
than the previous low-energy observations, and show that the 
variability time scale 
is indeed as short as a few hundred seconds even in the soft X-rays.  
As a result the power spectrum of soft X-rays is now extended to  
higher frequencies between 0.03 Hz and 0.005 Hz.  The present 
observations also appeared to have reached the true Poisson level 
of fluctuations in the source. 

The slope of the power-spectrum observed in the present observations 
is consistent with that observed at low energies with $EXOSAT$. 
Based on $EXOSAT$ low energy observations Papadakis \& Lawrence (1995) 
had also reported the presence of an excess in the power spectrum 
near 0.0004 Hz and modelled it as a broad Gaussian 
component in the power spectrum.  This broad component indicated the
presence of {\it quasi-periodic oscillation (QPO)} in the source.
A broad feature of excess power can also be seen in the present 
observations (Fig. 10 lower panel)  
between 0.0006 Hz and 0.0018 Hz, and has also been modelled by adding
a Gaussian (centred at 0.0011$\pm$0.0004 Hz) to the power-law.  
The addition of a  Gaussian did not 
change the slope of the power-law and led to an improvement in the
fit to the power spectrum.  The QPO frequency is 1.75 to 3.75 times higher 
than that observed with $EXOSAT$. However, since the dynamic range of 
low frequencies in the present observations is smaller than that 
of the $EXOSAT$ observations, the QPO frequency is not well determined.

Two models that can lead to a power-law shape of the power spectrum 
and which were widely discussed by  Papadakis \& Lawrence (1995)
are -- a) shot-noise model and b) hot-spot model. 
In a shot-noise model superposition of flares with the right shape 
can produce a power-law shape of the spectrum.  In a variation of this
model by  Begelman \& de Kool (1991), known as the `reservoir model',
a power-law shape with an index of -1.5 results from exponentially decaying
flares having a decay-time proportional to their amplitude, and their 
occurrence time is proportional to the amplitude of the previous flare. 
The resulting slope of the power-law is much flatter than the observed 
index in soft X-rays.
In a shot-noise model (Abramowicz et al. 1991; Bao \& Ostgaard 1994; 1995) 
the innermost parts of the accretion disk can develop orbiting 
hot-spots or clumps due to various thermal and viscous instabilities 
which modulate the intensity.  Bao \& Ostgaard (1995) have investigated 
power spectra of variability produced by orbiting blobs around black-holes
using full relativisitic details. They have studied the effects of intrinsic
luminosity of the blobs, their orbital radii, their number distribution,
the inclination of the accretion disk, and the 
properties of the accretion disk -- optically thick or thin.
The power-spectrum in every case is characterised by a power-law shape
(index -0.6 to -1.5) which flattens at the low frequency end and drops 
rapidly at the high frequency end.  The spectral slope is most
sensitive to the number distribution of blobs as a function of radius.
It is also sensitive to the inclination of the disk due to gravitational
lensing effect -- flattening at very high inclinations ($\geq$70\degrees).
None of the slopes is, however, as steep as found in the present 
$ROSAT$ observations or the previous $EXOSAT$ observations.  
A finer tuning of the blob models of  Bao \& Ostgaard
(1995) for the innermost regions of the accretion disk that are
responsible for the shape of the power-spectrum at the high frequency end 
might help in explaining the results.

A QPO-like feature has also been predicted by Bao \& Ostergaard (1994)
due to rotating spots in the accretion disk closest to the central 
object and when such a disk is viewed at moderate angles.  They predict
the centroid frequency of the QPO to be 
$\sim$4$\times$10$^{-3}$(R/R$_s$)$^{-1.5}$(M/10$^6$M$_{\odot})^{-1}$ Hz,
where R is the distance of the innermost spots from the SMBH, R$_s$
is the Schwarzschild radius, and M is the mass of the SMBH.  If the X-ray
QPO emission comes from outside 3R$_s$, then the mass of the SMBH should 
be in the range of 5--11$\times$10$^5$M$_{\odot}$, which is a few times
smaller than that implied by the fastest observed variability.  QPOs have
earlier been observed  in many observations of an AGN in NGC 5548 
(centroid frequency=0.002 Hz; Papadakis \& Lawrence 1993), and 
in several galactic black-hole candidates.

\section {CONCLUSIONS}

\begin{itemize}

\item [(i)] {Extended X-ray emission has been detected in the disks of 
NGC~1365 and NGC~4051, out to a radius of 2 kpc from their centres, and 
is much larger in extent than the emission line gas.}

\item [(ii)] {The  extended X-ray emission in NGC~1365 appears to be 
co-spatial with the starburst regions.
An elongated structure of X-ray emission breaking out of
the disk in NGC~1365 suggests the presence of superwinds from a starburst
as partly responsible for the extended emission in this galaxy.}

\item [(iii)] {The extended component of soft X-ray emission seems to dominate
the flux in NGC~4051 during its low intensity and low variability state. 
Nuclear activity as well as starburst induced activity could be responsible
for this component.} 

\item [(iv)] {The power spectrum of the rapid soft X-ray variability
in NGC~4051 has been extended to higher frequencies than before.   
It is best characterised by a constant + power-law + a Gaussian. 
The power-law slope is consistent with previous low-energy observations 
with $EXOSAT$. The presence of a Gaussian feature signifies the
persistence of a QPO seen earlier with $EXOSAT$.}

Spatially resolved spectral observations with $AXAF$ would be required
to fully understand the extended emission in these galaxies.

\end{itemize}

\acknowledgements
I wish to thank the entire $ROSAT$ team for making these observations 
possible.
I thank Dr. D.E. Harris for providing me with the software and scripts
for running the de-wobbling procedure on the $ROSAT$ HRI data.
I thank Drs. H. Kristen and P.O. Lindblad for providing their
published  optical maps of NGC~1365, and Dr. A.J. Holloway for providing 
the optical maps of NGC~4051.  I thank an anonymous referee for
his comments and suggestions leading to improvements in this paper.
I thank Dr. A.R. Rao for going through an earlier version of 
this manuscript and giving his useful comments.
This research has made use of the PROS software package provided
by the ROSAT  Science Data Center at Smithsonian Astrophysical
Observatory, and  FTOOLS software package provided  
by the High Energy Astrophysics Science Archive Research Center 
(HEASARC) of NASA's Goddard Space Flight Center.

\clearpage
\setcounter{page}{18}
\begin{figure}

\figurenum{1(a)}
\caption
{The corrected X-ray image of NGC~4051 as observed with the
$ROSAT$ HRI and smoothed with a Gaussian ($\sigma$=2\arcsec).   
Contours are plotted at 0.1, 0.15, 0.2, 0.3, 0.5, 1, 2, 5, 10, 20, 30, 
50, 70, 90 and 95\% of the peak brightness of 30.8 counts pixel$^{-1}$.
(1 pixel = 0.5\arcsec$\times$0.5\arcsec and 1\arcsec=66.9 pc)}
\end{figure}

\begin{figure}
\figurenum{1(b)}
\caption
{The corrected X-ray image of NGC~4051 as observed with the
$ROSAT$ HRI and smoothed with a Gaussian ($\sigma$=5\arcsec).   
Contours are plotted at 0.1, 0.15, 0.2, 0.3, 0.5, 1, 2, 5, 10, 20, 30, 
50, 70, 90 and 95\% of the peak brightness of 11.6 counts pixel$^{-1}$.
(1 pixel = 0.5\arcsec$\times$0.5\arcsec and 1\arcsec=66.9 pc)}
\end{figure}

\begin{figure}
\figurenum{1(c)}
\caption
{The $uncorrected$ X-ray image of NGC~4051 as observed with the
$ROSAT$ HRI and smoothed with a Gaussian ($\sigma$=2\arcsec).   
Contours are plotted at 0.1, 0.15, 0.2, 0.3, 0.5, 1, 2, 5, 10, 20, 30, 
50, 70, 90 and 95\% of the peak brightness of 27.2 counts pixel$^{-1}$.
(1 pixel = 0.5\arcsec$\times$0.5\arcsec and 1\arcsec=66.9 pc)}
\end{figure}

\begin{figure}
\figurenum{2(a)}
\caption
{The corrected X-ray image of NGC~1365 as observed with the
$ROSAT$ HRI and smoothed with a Gaussian ($\sigma$=2\arcsec).   
Contours are plotted at 5, 8, 12, 20, 30,
50, 70, 90 and 95\% of the peak brightness of 0.462 counts pixel$^{-1}$.
(1 pixel = 0.5\arcsec$\times$0.5\arcsec and 1\arcsec=90.2 pc)}
\end{figure}

\begin{figure}
\figurenum{2(b)}
\caption
{The corrected X-ray image of NGC~1365 as observed with the
$ROSAT$ HRI and smoothed with a Gaussian ($\sigma$=5\arcsec).   
Contours are plotted at 4, 6, 10, 20, 30,
50, 70, 90 and 95\% of the peak brightness of 0.221 counts pixel$^{-1}$.
(1 pixel = 0.5\arcsec$\times$0.5\arcsec and 1\arcsec=90.2 pc)}
\end{figure}

\begin{figure}
\figurenum{3}
\caption
{The azimuthally averaged radial brightness distribution of the
nuclear X-ray source in NGC~1365 extracted from the corrected
image shown in Fig.1.   The fit with the $ROSAT$ PSF is shown as a solid
curve.}
\end{figure}

%

\begin{figure}
\figurenum{4}
\caption
{The azimuthally averaged radial brightness distribution of the
nuclear X-ray source in NGC~4051 extracted from the corrected
image shown in Fig.2. The fit with the $ROSAT$ PSF is shown as a solid
curve.  }
\end{figure}

\begin{figure}
\figurenum{5}
\caption
{Contours of the X-ray source from Fig. 2(a)  overlaid on the B-band 
image (gray scale) of  NGC~1365 obtained from Kristen et al. (1997).}
\end{figure}

\begin{figure}
\figurenum{6}
\caption
{Contours of the X-ray source from Fig. 1(a) overlaid on the V-band image 
of NGC~4051 obtained from Christopoulou et al. (1997).}
\end{figure}

\begin{figure}
\figurenum{7}
\caption
{Contours of the extended component of X-ray source, obtained by 
subtracting a model image of a point source and smoothed with a 
Gaussian ($\sigma$=2\arcsec), overlaid on the V-band image of 
NGC~4051 obtained from Christopoulou et al. (1997).
The contour values are 0.1, 0.2, 0.5, 1, 2, 5, 10\% of the peak brightness 
shown in Fig. 1(a). }
\end{figure}

\begin{figure}
\figurenum{8}
\caption
{Contours of the X-ray source  Fig. 1(a) overlaid on the [O III] image 
of  NGC~4051 obtained from Christopoulou et al. (1997).  }
\end{figure}

\begin{figure}
\figurenum{9}
\caption
{$ROSAT$ HRI light curve of  NGC~4051.   }
\end{figure}

\begin{figure}
\figurenum{10}
\caption
{Power spectral density of the short-term variability in  NGC~4051.  
The upper panel shows the power spectrum based on data binned every 16s.
The best fit constant+power-law model is shown as a solid curve.  The lower 
panel shows the power spectrum based on data binned every 32s.  The best 
fit constant+power-law model is shown as a dotted curve, and the best fit
constant+power-law+Gaussian model is shown as a solid curve. }
\end{figure}

\end{document}


\setcounter{page}{16}

\begin{planotable}{lccccccc}
\tablenum{1}
\tablecaption{Basic Parameters of the two Seyferts}
\tablewidth{45pc}
\tablehead{
\colhead{Name} &
\colhead{Distance} &
\colhead{Magnitude} &
\colhead{Log L$_{\rm Hx}$}  & 
\colhead{X-ray Size}  &  
\colhead{N$_{\rm H}$} & 
\colhead{F$_{\rm Sx}^a$} & 
\colhead{Log L$_{\rm Sx}^b$}  \\  
\colhead{} &
\colhead{Mpc} &
\colhead{B} & 
\colhead{ergs s$^{-1}$} &
\colhead{arcsecs} &
\colhead{10$^{20}$cm$^{-2}$} &
\colhead{10$^{-11}$ergs cm$^{-2}$s$^{-1}$} &
\colhead{ergs s$^{-1}$} 
}
\startdata
NGC 1365 &  18.6$\pm$1.9 & 9.45 & 40.6$^c$     & 6--22 &  1.35 & 0.11  & 40.65 \\ 
NGC 4051 &  13.8$^d$   & 10.6 & 41.2--41.9$^e$ & 6--38 &  1.3  & 4.75   & 42.03 \nl
\enddata
\tablenotetext{a}{Unabsorbed total source flux in the $ROSAT$ HRI band of 0.1--2.0 keV; present work (\S4.1).}
\tablenotetext{b}{Total source luminosity in the 0.1--2.0 keV energy band; present work. (\S4.1)}
\tablenotetext{c}{2--10 keV luminosity from $ASCA$  measurements by Iyomoto et al. 1997.}
\tablenotetext{d}{Redshift=0.0023; Hubble constant= 50 km s$^{-1}$ Mpc$^{-1}$.}
\tablenotetext{e}{2--10 keV luminosity from $Ginga$ measurements by Matsuoka et al. 1990.}

\end{planotable}

\begin{planotable}{llcccc}
\tablenum{2}
\tablecaption{Details of ROSAT HRI observations}
\tablewidth{45pc}
\tablehead{
\colhead{Name} &
\colhead{Sequence No.} &
\colhead{Start Time} &
\colhead{End Time} &
\colhead{Effective Exposure} &
\colhead{Mean Count Rate$^a$} \\
\colhead{} &
\colhead{} &
\colhead{Y, M, D, UT} &
\colhead{Y, M, D, UT} &
\colhead{Time (s)} &
\colhead{counts s$^{-1}$}
}
\startdata
NGC 1365 & rh701297n00 & 1994 07 20 21:15:21 & 1994 08 04 18:15:36  & 9826 & 0.023$\pm$0.002  \\
NGC 1365 & rh701297a01 & 1994 08 23 12:43:48 & 1994 08 23 12:49:42  & 346  &  0.018$\pm$0.009 \\ 
NGC 1365 & rh701297a02 & 1995 07 03 17:39:20 & 1995 07 04 19:25:12  & 9762 & 0.022$\pm$0.002  \\
NGC 4051 & rh701298n00 & 1994 05 21 07:04:57 & 1994 05 22 15:16:20  & 10478 & 1.08$\pm$0.01 \nl
\enddata
\tablenotetext{a}{Mean count rates are after background subtraction and 
integrated over the full extent of the source: point source + extended 
component.}
\end{planotable}

\begin{planotable}{ccccccc}
\tablenum{3}
\tablecaption{NEW X-RAY SOURCES DETECTED WITH THE ROSAT HRI}
\tablewidth{45pc}
\tablehead{
\colhead{S. No.} &
\colhead{R.A. (2000)} &
\colhead{Dec. (2000)} &
\colhead{Count Rate$^a$} &
\colhead{SNR$^b$} &
\colhead{Years of} &
\colhead{ID$^c$} \\
\colhead{} &
\colhead{H~~M~~S} &
\colhead{D~~M~~S} &
\colhead{} &
\colhead{} &
\colhead{Detection} &
\colhead{}
}
\startdata
 1. & 03~33~34.4 & -36~09~36.2  & 2.94$\pm$0.66 & 4.48 & 1994   & A Transient in \\ 
    &            &              &               &      &      & inner arm of NGC~1365 \\
 2. & 03~33~11.4 & -36~11~34.6  & 2.54$\pm$0.60 & 4.26 & 1994, 1995 & 2 objects (B$>$22,  B=20.8) \\ 
    &            &              &               &      &      & within 6\arcsec \\

 3. & 03~33~12.8 & -36~11~51.7  & 2.12$\pm$0.55 & 3.85 & 1994, 1995 & B=19.8 mag object\\ 
    &            &              &               &      &      & within 2.6\arcsec \\
 4. & 03~32~53.3 & -36~06~44.4  & 1.66$\pm$0.49 & 3.39 & 1994  & B=22.9 mag object \\ 
    &            &              &               &      &      & within 7\arcsec \\
 5$^d$. & 03~33~12.1 & -36~19~43.4  & 18.5$\pm$1.50 & 12.3 & 1994, 1995 & 1E0331.3-3629, B=20.4 \\ 
    &            &              &               &      &      & object within 4.5\arcsec \\
 6$^d$. & 03~33~10.8 & -36~19~50.5  & 13.0$\pm$1.30 & 10.1 & 1994, 1995 & 1E0331.3-3629, B$\geq$23.5 \\
    &            &              &               &      &      & object within 4\arcsec \\
 7. & 03~34~07.6 & -36~03~59.3  & 3.44$\pm$0.67 & 5.12 & 1994, 1995 & EQ0332-3614, B$\geq$24.0 \\
    &            &              &               &      &      & object within 5\arcsec \\
 8. & 03~32~36.7 & -36~11~11.9  & 2.0$\pm$0.90  & 2.5  & 1994  & B=23.5 object \\
    &            &              &               &      &      & within 12\arcsec \\
 9. & 03~33~15.9 & -36~18~13.0  & 2.70$\pm$0.90 & 3.0  & 1995  & B=20.9 object \\
    &            &              &               &      &      & within 5\arcsec \\
10. & 03~33~40.2 & -36~13~38.0  & 1.80$\pm$0.60 & 3.0 &  1995  & No visible counterpart \\ 
    &            &              &               &      &      & within 20\arcsec \\
11. & 03~33~41.9 & -36~07~33.0  & 2.00$\pm$0.60 & 3.5 &  1995  & Big bright patch in NGC~1365\\
    &            &              &               &      &      & H II region or a cluster \\
12$^e$. & 03~32~22.6 & -36~06~33.0  & 4.00$\pm$1.00 & 4.0 &  1994  & B=23 mag object  \\ 
    &            &              &               &      &      & 31\arcsec away \\
13$^e$. & 03~33~48.1 & -36~25~18.0  & 7.00$\pm$2.00 & 3.5 & 1995    & B=19.3 mag object \\
    &            &              &               &      &      & within 11\arcsec \\
14. & 12~03~58.9 & +44~37~23.9  & 1.69$\pm$0.57 & 3.0 & 1994 & B=20 mag object \\
    &            &              &               &      &      & within 9\arcsec \\
15. & 12~02~32.4 & +44~39~45.0  & 2.90$\pm$0.80 & 3.5 & 1994 & no visible counterpart \\
    &            &              &               &      &      & within 35\arcsec \\
16. & 12~04~13.8 & +44~31~52.0  & 2.20$\pm$0.90 & 2.5 & 1994 & B=19.8 mag object \\
    &            &              &               &      &      & within 5\arcsec \nl
\enddata
\tablenotetext{a}{in units of 10$^{-3}$ s$^{-1}$.}
\tablenotetext{b}{SNR: Signal-to-Noise Ratio.}
\tablenotetext{c}{B magnitudes and distances of candidate objects from 
source position are given; EQ refers to clusters of galaxies.}
\tablenotetext{d}{Source Numbers 5 and 6 may be part of one extended source.}
\tablenotetext{e}{Close to the edge of the HRI field of view.}
\end{planotable}